\documentclass[twocolumn,secnumarabic,amssymb, nobibnotes, aps, pra,superscriptaddress]{revtex4-2}

\usepackage{etoolbox}
\usepackage{graphicx}  
\usepackage{epstopdf}
\usepackage{dcolumn}   
\usepackage{bm}        
\usepackage{amssymb}   
\usepackage{amsmath}   
\usepackage{amsthm}
\usepackage{chemarrow}
\usepackage{color}
\usepackage{mathrsfs}
\usepackage{float}
\usepackage{bbold}
\usepackage{bbm}

\begin{document}


\title{Protection of quantum evolutions under parity-time symmetric non-Hermitian Hamiltonians  by dynamical decoupling}

\date{\today}

\author{Ji Bian}
\thanks{These authors contributed equally to this work.}
\affiliation{School of Physics and Astronomy,
	Sun Yat-Sen University,
	Zhuhai,
	519082,
	China}
\affiliation{Center of Quantum Information Technology,
	Shenzhen Research
		Institute of Sun Yat-sen University,
	Shenzhen,
	518087,
China}
\author{Kunxu Wang}
\thanks{These authors contributed equally to this work.}
\affiliation{School of Physics and Astronomy,
	Sun Yat-Sen University,
	Zhuhai,
	519082,
	China}
\author{Pengfei Lu}
\affiliation{School of Physics and Astronomy,
	Sun Yat-Sen University,
	Zhuhai,
	519082,
	China}
\author{Xinxin Rao} 
\affiliation{School of Physics and Astronomy,
	Sun Yat-Sen University,
	Zhuhai,
	519082,
	China}
\author{Hao Wu}
\affiliation{School of Physics and Astronomy,
	Sun Yat-Sen University,
	Zhuhai,
	519082,
	China}
\author{Qifeng Lao}
\affiliation{School of Physics and Astronomy,
	Sun Yat-Sen University,
	Zhuhai,
	519082,
	China}
\author{Teng Liu}
\affiliation{School of Physics and Astronomy,
	Sun Yat-Sen University,
	Zhuhai,
	519082,
	China}
\author{Yang Liu}
\affiliation{School of Physics and Astronomy,
	Sun Yat-Sen University,
	Zhuhai,
	519082,
	China}
\affiliation{Center of Quantum Information Technology,
	Shenzhen Research
	Institute of Sun Yat-sen University,
	Shenzhen,
	518087,
	China}
\author{Feng Zhu}
\affiliation{School of Physics and Astronomy,
	Sun Yat-Sen University,
	Zhuhai,
	519082,
	China}
\affiliation{Center of Quantum Information Technology,
	Shenzhen Research
	Institute of Sun Yat-sen University,
	Shenzhen,
	518087,
	China}
\author{Le Luo}
\email{luole5@mail.sysu.edu.cn}
\affiliation{School of Physics and Astronomy,
	Sun Yat-Sen University,
	Zhuhai,
	519082,
	China}
\affiliation{Center of Quantum Information Technology,
	Shenzhen Research
	Institute of Sun Yat-sen University,
	Shenzhen,
	518087,
	China}
\affiliation{
State Key Laboratory of Optoelectronic
		Materials and Technologies,
Sun Yat-Sen
		University (Guangzhou Campus),
Guangzhou,
510275,
China}

\begin{abstract}
Parity-time ($\mathcal{PT}$) symmetric non-Hermitian Hamiltonians bring about many novel features and interesting applications such as quantum gates faster than those in Hermitian systems, and topological state transfer. The performance of evolutions under $\mathcal{PT}$-symmetric Hamiltonians is degraded by the inevitable noise and errors due to system-environment interaction and experimental imperfections. In contrast to Hermitian Hamiltonians, the fluctuations in dissipative beams that are utilized to generate non-Hermitian contributions in the $\mathcal{PT}$-symmetric Hamiltonians cause additional errors. Here we achieve the protection of $\mathcal{PT}$-symmetric Hamiltonians against noise acting along the
qubit's quantization axis by combining quantum evolutions with dynamical decoupling sequences. We demonstrate the performance of our method by numerical simulations. Realistic noise sources and parameters are chosen including: constant detuning error, time-varying detuning noise and dissipative-beam noise. The fidelities of the protected evolutions are well above the unprotected ones under all the above situations.
Our work paves the way for further studies and applications of non-Hermitian $\mathcal{PT}$-symmetric physics in noisy quantum systems.
\end{abstract}
\maketitle
\section{Introduction}
In quantum mechanics, the Hermiticity requirement of  Hamiltonians guarantees the energy of a system to be real. However, as demonstrated in
\cite{bender1999pt}, a class of non-Hermitian Hamiltonians
satisfying parity-time ($\mathcal{PT}$) symmetry can still exhibit real
eigenenergies. $\mathcal{PT}$-symmetric Hamiltonians exhibit various exotic
behaviors, in which a key property is $\mathcal{PT}$-symmetry-breaking
transitions that occur at an exceptional point (EP). EP is a point in
the parameter space where the eigenvalues and eigenstates of the Hamiltonian
coalesce
\cite{bender2007faster,zheng2013observation,li2019observation,wu2019observation,wang2021observation,ding2021experimental,biana2022quantum}.
Exotic properties of $\mathcal{PT}$-symmetric systems have been experimentally studied in various classical systems \cite{feng2014single,assawaworrarit2017robust,alaeian2014parity,zhu2014p,bittner2012p}.
This has stimulated many applications such as unidirectional
light transport \cite{feng2011nonreciprocal}, single-mode lasers
\cite{hodaei2014parity}, optimal energy transfer
\cite{doppler2016dynamically} and enhanced sensing
\cite{hodaei2017enhanced}. 
Recently, $\mathcal{PT}$-symmetric Hamiltonians are
also constructed in genuine quantum systems, e.g., ultracold atoms
\cite{li2019observation}, NV-centers \cite{wu2019observation},
trapped ions \cite{wang2021observation,ding2021experimental}, and
superconducting quantum circuits \cite{naghiloo2019quantum}. These
allow quantum signatures such as perfect quantum coherence at EP to
be revealed \cite{wang2021observation}. The novel character of $\mathcal{PT}$-symmetric Hamiltonians are also expected to play an important role in quantum control. For example, faster-than-Hermitian quantum mechanics evolutions could be achieved \cite{zheng2013observation,bender2007faster}; topological
structure of exceptional points could be utilized to realize robust
quantum control \cite{liu2021dynamically}.

In order to study and utilize $\mathcal{PT}$-symmetric properties in quantum systems, one needs to construct general quantum evolutions ($U_{\textrm{PT}}$) under $\mathcal{PT}$-symmetric Hamiltonians ($H_{\textrm{PT}}$), e.g., a faster-than-Hermitian quantum mechanics evolution \cite{zheng2013observation,bender2007faster}, or encircling an exceptional point \cite{liu2021dynamically}. As the unavoidable environmental perturbations degrade the performance of quantum gates and evolutions, protecting the system against noise and errors is key to further studies. Apart from the noise sources such as magnetic field noise and detuning errors that are common to Hermitian systems, additional noise sources present in $\mathcal{PT}$-symmetric quantum systems. For example, a natural and popular way to construct $\mathcal{PT}$-symmetric Hamiltonian is through the application of a dissipative beam \cite{li2019observation,ding2021experimental}. However, the fluctuations and errors in dissipative beams usually become the dominant error sources in these experiments \cite{ding2021experimental}. Developing methods to mitigate the above noise and errors is thus a vital step towards exploring $\mathcal{PT}$-symmetric physics in quantum systems, but so far related studies is still lacking.

Here we demonstrate the protection of $U_{\textrm{PT}}$ by dynamical decoupling (DD) \cite{viola1999dynamical,suter2016colloquium,souza2012robust,yang2011preserving,degen2017quantum,biercuk2011dynamical}. 
DD is a popular method to protect the quantum system against static or time-dependent environmental noise \cite{suter2016colloquium,yang2011preserving}. By applying a series of control operations to the quantum system, the unwanted system-environment interaction could be cancelled to a large extent.
DD has been utilized to protect both quantum memories (null operations) \cite{yang2011preserving,wang2017single,souza2012robust,peng2011high,biercuk2009experimental} and quantum gates (arbitrary evolutions) \cite{zhang2014protected,khodjasteh2009dynamically,bermudez2012robust,xu2012coherence,souza2015high,ng2011combining,khodjasteh2010arbitrarily,piltz2013protecting}, without additional qubits. 
In this work, we first demonstrate how to modify  general quantum evolutions under $\mathcal{PT}$-symmetric Hamiltonians in such a way that they can be combined with DD operations, thus achieving the desired evolution goal under DD protection. We then present numerical simulations of general $U_{\textrm{PT}}$ under realistic noise sources including:
constant detuning error, time-varying detuning noise and importantly, dissipation beam noise. The performance of DD-protected evolutions is significantly better than unprotected ones in all these situations. To our knowledge, this is  the first time that dynamical decoupling bridges to the non-Hermitian Hamiltonian. We not only demonstrate for the first time that $\mathcal{PT}$-symmetric Hamiltonians can be protected against environmental noises, but also provide a novel DD scheme to realize such protection. Experiments using our method for trapped ion qubits is currently being implemented in our lab. This work pioneer a new research route,  enabling studies and applications of non-Hermitian $\mathcal{PT}$-symmetric physics with noisy quantum systems.


\section{Construction of $\mathcal{PT}$-symmetric Hamiltonians}\label{sec2}

\begin{figure}[h]  
	\makeatletter
	\def\@captype{figure}
	\makeatother
	\includegraphics[scale=1]{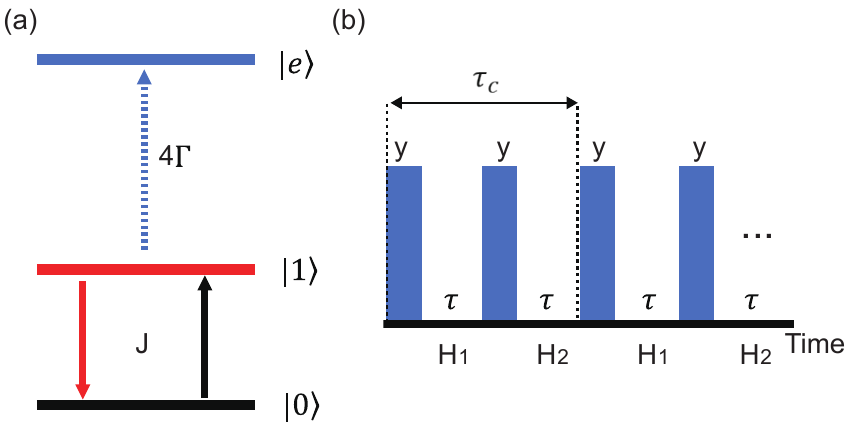}
	\caption{(a) Construction of $\mathcal{PT}$-symmetric Hamiltonians in qubit systems by the dissipative scheme. A control field with strength $J$ couples two qubit levels. The population on state $|1\rangle$ is pumped to other levels outside the qubit space, this could be achieved by a dissipative beam plus spontaneous emission. A population loss on state $|1\rangle$  is thus produced, with loss rate $4\Gamma$ which is controlled by the dissipative beam strength. (b) DD-protected evolutions under $\tilde{H}_{\textrm{PT}}$. The whole protection operation is established by repeating the basic unit with duration $\tau_c$. The rectangles represent $\pi$ rotations, and $H_{k}$ is the Hamiltonian in between.}
	\label{fig0}
\end{figure}

The $\mathcal{PT}$-symmetric Hamiltonian considerded here reads
\begin{equation}\label{h1}
H_{\textrm{PT}}=2i\Gamma I_z+2J I_x,
\end{equation}
where
\begin{equation*}
I_x= \frac{1}{2}\begin{pmatrix} 0 & 1 \cr 1 & 0 \end{pmatrix}, \
I_y= \frac{1}{2}\begin{pmatrix} 0 & -i \cr i & 0 \end{pmatrix}, \
I_z= \frac{1}{2}\begin{pmatrix} 1 & 0 \cr 0 & -1 \end{pmatrix}
\end{equation*}
are angular momentum operators and $\Gamma, J$ are real
parameters. Indeed, $[\textrm{PT},H_{\textrm{PT}}]=0$, satsifying
the $\mathcal{PT}$ requirement, where $\textrm{P}=2I_x$,
$\textrm{T}=*$ denotes complex conjugation operation.
The exceptional point is located at $\Gamma=J$. When $\Gamma<J$, the system is in the $\mathcal{PT}$-symmetry preserving phase, and when $\Gamma>J$, it is in the $\mathcal{PT}$-symmetry broken phase.
This Hamiltonian has not been directly implemented in quantum systems due to the difficulties in realizing the gain on $|0\rangle$.  It has been constructed by e.g., a dilation method using two qubits \cite{wu2019observation}. Alternatively, a popular way to study $H_{\textrm{PT}}$ is to use a dissipative scheme that generates a state-dependent loss in a single qubit, 
as is done in, e.g., cold atoms \cite{li2019observation}, superconducting circuits \cite{naghiloo2019quantum}, and trapped ions \cite{wang2021observation,ding2021experimental}.
In these experiments, the qubit levels $|0\rangle$ and $|1\rangle$ are coupled by a control field (e.g., a microwave field)
with coupling strength $J$, and a loss of population on 
$|1\rangle$ with effective loss rate $4\Gamma$ is constructed, here $|0\rangle$ ($|1\rangle$) are eigenstates of $I_z$ with eigenvalues $\frac{1}{2}$($-\frac{1}{2}$).
This will generate a passive $\mathcal{PT}$-symmetric Hamiltonian 
\begin{equation}\label{hptmap}
\tilde{H}_{\textrm{PT}}=2JI_x-2i\Gamma |1\rangle \langle 1|=2i\Gamma I_z+ 2JI_x-i\Gamma\mathbf{I},
\end{equation}
where $\mathbf{I}$ is the identity matrix and $\Gamma \geq 0$ because the dissipation rate cannot be made negative.
The state evolution $|\psi(t)\rangle$ under the original $\mathcal{PT}$-symmetric Hamiltonian $H_{\textrm{PT}}$ could be recovered by multiplying a factor $e^{\Gamma t}$ to the state $|\tilde{\psi}(t)\rangle$ evolving under $\tilde{H}_{\textrm{PT}}$. 
As the factor $e^{\Gamma t}$ does not affect the EP or the essential dynamics of the $\mathcal{PT}$-symmetric systems, it is irrelevant in typical studies on  $\mathcal{PT}$-symmetric physics \cite{zheng2013observation,naghiloo2019quantum,wen2020observation}, the protection of $U(\tilde{H}_{\textrm{PT}},T)$ is thus considered equivalent to the protection of $U(H_{\textrm{PT}},T)$ in this work.
The population loss 
could be achieved experimentally by adding a dissipation beam
\cite{ding2021experimental}, as illustrated in Fig.\ref{fig0}, (a). 
Note that the fluctuations and errors in dissipative beams usually become the dominant error sources \cite{ding2021experimental}, and is exclusive to non-Hermitian-Hamiltonian construction.

\section{Dynamical decoupling protected evolutions under $\mathcal{PT}$-symmetric Hamiltonians}\label{sec3}

The dynamics of the qubit in a general decohering
environment is governed in a semi-classical picture by a
Hamiltonian of the form \cite{biercuk2011dynamical}
\begin{small}
	\begin{equation}\label{hnoise}
	\begin{aligned}
	H&=\tilde{H}_{\textrm{PT}}+H_{\textrm{n}} \\
	&=[2i\Gamma+2i\delta_{\Gamma}(t)+2\beta(t)] I_z+ [2J+\alpha(t)]I_x-i[\Gamma+\delta_{\Gamma}(t)]\mathbf{I},
	\end{aligned}
	\end{equation}
\end{small}
where $H_{\textrm{n}}=[2i\delta_{\Gamma}(t)+2\beta(t)]I_z+\alpha(t)I_x-i\delta_{\Gamma}(t)\mathbf{I}$ is the noise Hamiltonian. $\alpha$, $\beta$ represent random fields imparted by the environment, and we explicitly use $\delta_{\Gamma}(t)$ to represent noise in dissipative beam. 
In the following we only consider noises coupled to $I_z$ and set $\alpha=0$ and $\beta, \delta_{\Gamma} \neq 0$.
$\beta$ and $\delta_{\Gamma}$ could also represent experimental imperfections and errors.

In order to protect the system from unwanted environmental noise, while still achieving the evolution under $\tilde{H}_{\textrm{PT}}$ for time $T$ ($U(\tilde{H}_{\textrm{PT}},T)$),  we can split the target evolution into pieces and insert them in the free precession intervals of a standard DD sequence. 
 
\subsection{Protection based on a CPMG-like sequence}\label{cpmglike}
 To demonstrate the main idea, we first choose a simple Carl-Purcell-Meiboom-Gill (CPMG)-like sequence  \cite{souza2012robust} as the basic DD sequence, and add the control Hamiltonian $H$ at the end of it in each cycle \cite{ng2011combining}, as demonstrated in Fig. \ref{fig0}, (b). 
 The evolution of the system for one cycle in this sequence is 
\begin{equation}
U=\prod_{k=1}^{N}{U(H_k,\tau_k) P_k},
\end{equation}
here  $\tau_k=\tau$, $N=2$, and $P_k$ are $\pi$ rotations along  $y$ axes.  $P_k$ are assumed to be instantaneous and perfect, thus the total duration of one cycle $\tau_c=2\tau$. 
$H_{1}=H_{\textrm{n}}$ which means we let the system evolves freely during this period, and the evolution is governed only by noise terms. $H_{2}=H$ meaning that the control pulses are presented during this period. The sequence is repeated $m$ times, with the effective evolution time (under $\tilde{H}_{\textrm{PT}}$)  $T=m\tau$.  
This sequence is simple in that the evolution under $H$ does not need to be split in one cycle, similar to the DD-protected gates studied in \cite{ng2011combining}. 
This sequence provides first order protection for $U(\tilde{H}_{\textrm{PT}},T)$ \cite{zhang2014protected}. To show this result, we use average Hamiltonian theory \cite{brinkmann2016introduction}. 

Average Hamiltonian theory states  that an evolution $U(\tau_c)$, such as a DD cycle under a time dependent Hamiltonian, can be described by an evolution governed by an effective time-independent (or average) Hamiltonian
\begin{equation}
\bar{H}=\bar{H}^{(1)}+\bar{H}^{(2)}+\bar{H}^{(3)}+...,
\end{equation}
where $\bar{H}^{(k)}$ represents $k$th order approximation \cite{zhang2014protected}. The explict expressions for $\bar{H}^{(k)}$ could be found in e.g., \cite{brinkmann2016introduction,ng2011combining} by calculating the Magnus expansion, and the first-order approximation is 
\begin{equation}
\bar{H}^{(1)}=\frac{1}{\tau_c} \int_{0}^{\tau_c}dt H(t).
\end{equation}
Assume $\tau$ and $m$ are chosen such that $\beta$ and $\delta_{\Gamma}$ are kept nearly constant in a single cycle. As $P_{1}=P_{2}=-P^{\dagger}_{1}$, the evolution $P_{2} U(H_{1},\tau) P_{1}$ is equivalent (omitting the irrelevant global phase) to 
\begin{equation}
U(H',\tau)=e^{-i H' \tau}=e^{-i P_1 H_{\textrm{1} } P^{\dagger}_1 \tau},
\end{equation} 
where 
\begin{equation}\label{hprime}
H'=P_1 H_{1} P^{\dagger}_1 
=-H_{\textrm{n}}-2i\delta_{\Gamma}\mathbf{I}.
\end{equation} 
The evolution for one cycle under first-order approximation is then
\begin{equation}\label{adding}
\begin{aligned}
U(\tau) &=U(H_{2},\tau)P_2U(H_{1},\tau)P_1 =U(H,\tau)U(H',\tau) \\
 &=e^{-i \bar{H} \tau_c}\approx e^{-i \bar{H}^{(1)} \tau_c}=e^{-i (\frac{\tau}{\tau_c}(H+H')) \tau_c}\\
 &=e^{-i \tilde{H}_{\textrm{PT}} \tau}e^{-2\delta_{\Gamma}\tau}.
\end{aligned}
\end{equation}
So the evolution for $m$ cycle is
\begin{equation}
U_m(\tau)=\prod^{m}{U(\tau)} \approx 
e^{-i \tilde{H}_{\textrm{PT}} T}e^{-2\tilde{\delta}_{\Gamma}T}.
\end{equation}
where $\tilde{\delta}_{\Gamma}$ is the average of $\delta_{\Gamma}$ over $m$ cycles.

If the dissipative beam noise is not presented, i.e., $\delta_{\Gamma}=0$, we have $U_m=U(\tilde{H}_{\textrm{PT}},T)$. That is, we achieve first order protection of the desired evolution $U(\tilde{H}_{\textrm{PT}},T)$. If $\delta_{\Gamma} \neq 0$, $U_m$ will differ from $U(\tilde{H}_{\textrm{PT}},T)$ by a factor $e^{-2\bar{\delta}_{\Gamma}T}$. In practice, this is not a concern (also discussed in Sec. \ref{sec2}), as the EP remains unchanged with or without this factor. The transition from $\mathcal{PT}$-symmetry preserving phase to $\mathcal{PT}$-symmetry broken phase, as well as the essential dynamics are not affected by this factor either \cite{ding2021experimental}.  So the sequences presented in this work are still considered to achieve the protection of $U(\tilde{H}_{\textrm{PT}},T)$ with the irrelevant factor $e^{-2\bar{\delta}_{\Gamma}T}$ ignored.

The above sequence achieves first-order protection, because the noise term $H_n$ in $H'$ changes sign, and cancels out $H_n$ in $H$,  when adding $H$ and $H'$ in \eqref{adding}. As it is more convenient to start with a sequence where $J$ and $\Gamma$ are applied simultaneously, plus the fact that $\Gamma \geq 0$ cannot be made negative, we design the sequence with evolutions under $H_n$ alone to achieve the cancellation \cite{ng2011combining}. Otherwise, the $2i\Gamma I_z$ term between two DD pulses will also change sign and cancel out with the following $2i\Gamma I_z$ term,  and the protection scheme is failed.

Variations of $\beta(t)$ and $\delta_{\Gamma}(t)$
between successive evolutions under $H_k$ in a single cycle reduces the fidelity of the operation at the conclusion of the sequence. Nevertheless, a high enough fidelity could still be achieved using sufficiently small $\tau$, as demonstrated by the numerical simulation in Sec. \ref{ns}.
Note that more robust DD sequences e.g., $XY-4$ or KDD sequence \cite{souza2012robust}, might also be utilized to achieve the protection. Here for simplicity we use the DD sequence with a single rotation axis to demonstrate the main idea.

\subsection{Protection based on the CPMG sequence}\label{cpmg}
\begin{figure}[h]  
	\makeatletter
	\def\@captype{figure}
	\makeatother
	\includegraphics[scale=1]{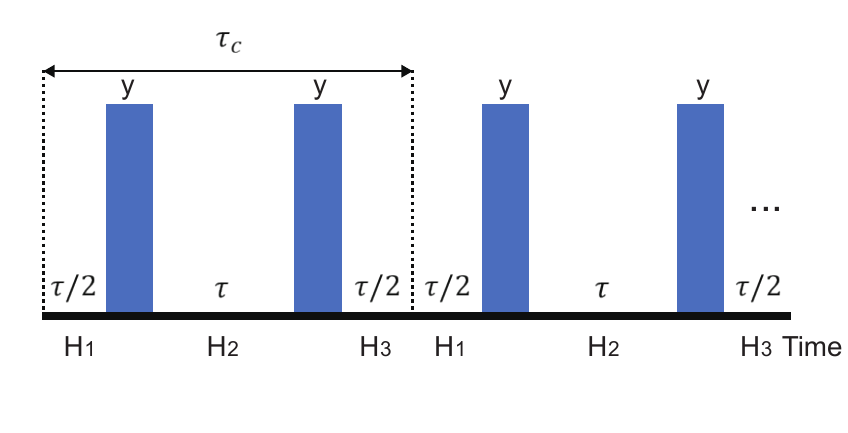}
	\caption{The protection based on the CPMG sequence. $\tau_c$ is the duration of a single cycle. }
	\label{figseq2}
\end{figure}
The sequence presented in the above subsection could be upgraded to achieve higher-order protection. It is proved that if $H(t)$ is symmetric over one cycle, i.e.,  $H(t)=H(\tau_c-t)$ $(0 \leq t \leq \tau_c)$, then all even order average Hamiltonian terms are equal to zero \cite{brinkmann2016introduction,ng2011combining}. This allows us to design a protection scheme based on the exact CPMG sequence \cite{souza2012robust} as shown in Fig. \ref{figseq2}, where $N=3$, $P_1=\mathbf{I}$, $P_{2,3}$ are $\pi$ rotations along  $y$ axes, $H_1=H_3=H$, $H_2=H_n$, and $\tau_1=\tau_3=\tau_2/2=\tau/2$. The evolution operator for one cycle up to the second order approximation is then (omitting the irrelevant global phase)
\begin{equation}\label{adding2}
\begin{aligned}
U &=U(H_{3},\tau/2)P_3U(H_{2},\tau)P_2U(H_{1},\tau/2)\\ 
&=U(H,\tau/2)U(H',\tau)U(H,\tau/2) \\
&=e^{-i \bar{H} \tau_c}\approx e^{-i (\bar{H}^{(1)}+\bar{H}^{(2)}) \tau_c},
\end{aligned}
\end{equation}
where $H'$ is the same as \eqref{hprime}. As
\begin{equation*}
\bar{H}^{(1)}=\frac{1}{\tau_c}(2H\tau/2+H'\tau)=\frac{1}{2}(\tilde{H}_{\textrm{PT}}-2i\delta_{\Gamma}\mathbf{I}),
\end{equation*}
and \cite{brinkmann2016introduction}
\begin{equation*}
\begin{aligned}
\bar{H}^{(2)}&=\frac{1}{2i\tau_c}\{[H',H_1]\frac{\tau^2}{2}+[H_3,H_1]\frac{\tau^2}{4}+[H_3,H']\frac{\tau^2}{2}\}\\
&=0,
\end{aligned}
\end{equation*}
the evolution for one cycle up to second order approximation is then
\begin{equation*}
U \approx e^{-i \tilde{H}_{\textrm{PT}} \tau}e^{-2\delta_{\Gamma}\tau},
\end{equation*}
and $U_m \approx e^{-i \tilde{H}_{\textrm{PT}} T}e^{-2\tilde{\delta}_{\Gamma}T}$. This sequence thus achieves second-order protection (actually all $\bar{H}^{(k)}$ with even $k$ equal $0$), and is more effective than the one in Fig. \ref{fig0}, (b). It is more complicated in that $H$ is split and placed before and after the DD sequence in one cycle. More advanced sequences, e.g., the concatenated DD (CDD) and the non-equidistant Uhrig DD (UDD)  can achieve arbitrary-order protections for a ``no operation" \cite{souza2012robust}. CDD-like sequences are further utilized to achieve the protection of arbitrary gates to any given order \cite{khodjasteh2010arbitrarily}, but the ability to realize the inverse of the gate operation (without affecting the noise terms during the gate operation) is required, and are thus not directly applicable in protecting $\tilde{H}_{\textrm{PT}}$ where $\Gamma$ cannot be negative. This interesting difference between Hermitian and non-Hermitian Hamiltonian protection, as well as the higher-order protection of $\tilde{H}_{\textrm{PT}}$ are worth further investigation.

\section{simulation results}\label{ns}
Simulation results are presented here to demonstrate the performance of our method.  The parameters in \eqref{hptmap} are chosen according to typical trapped  $^{171}$Yb$^{+}$ ion setups \cite{olmschenk2007manipulation,ding2021experimental}, which are promising platforms for future studies on $\mathcal{PT}$ symmetry. The present simulation could be adapted to other platforms by resetting these parameters.
As typically only the normalized density matrices and state fidelities are considered, e.g.,
when studying the faster-than-Hermitian quantum mechanics evolution \cite{zheng2013observation}, the $\mathcal{PT}$-symmetric phase transition behavior \cite{naghiloo2019quantum}, and the information flow \cite{wen2020observation}, we choose the normalized state fidelity  \cite{wen2020observation, suter2016colloquium}
\begin{equation}\label{fid}
F=\frac{\left| \mathrm{Tr}[ \rho _{\textrm{id}}(T)\rho_m(\tau) ] \right|}{\sqrt{\mathrm{Tr}\left( \rho _{\textrm{id}}(T)\rho _{\textrm{id}}(T) \right) \mathrm{Tr}[ \rho_m(\tau) \rho_m(\tau) ]}},
\end{equation}
where
\begin{equation*} \rho_{\textrm{id}}(T)=\frac{U(\tilde{H}_{\textrm{PT}},T)|\psi(t=0)\rangle \langle \psi(t=0)|U^{\dagger}(\tilde{H}_{\textrm{PT}},T)}{\textrm{Tr}[U(\tilde{H}_{\textrm{PT}},T)|\psi(t=0)\rangle \langle \psi(t=0)|U^{\dagger}(\tilde{H}_{\textrm{PT}},T)]}
\end{equation*} 
is the ideal density matrix with initial state $|\psi(t=0)\rangle$ and evolution time $T$, 
\begin{equation*}
\rho_m(\tau)=\frac{U_m(\tau)|\psi(t=0)\rangle \langle \psi(t=0)|U^{\dagger}_m(\tau)}{\textrm{Tr}[U_m(\tau)|\psi(t=0)\rangle \langle \psi(t=0)|U^{\dagger}_m(\tau)]}
\end{equation*}
is the actual density matrix with noise after the $m$ cycle sequence, to characterize the performance of the protection scheme. 

In the following, we show the protection of arbitrary evolutions under \eqref{hptmap} (or \eqref{h1} as they generate the same normalized density matrix), and in particular, the faster-than-Hermitian quantum mechanics non-Hermitian NOT gate
$U_{\textrm{NOT}}=e^{-iH_\textrm{PT}T_\textrm{not}}$,
where $J>\Gamma$, the initial state is $|\psi(t=0)\rangle=|1\rangle$, and $T_{\textrm{not}}=\frac{\pi -2\!\:\mathrm{arcsin(}\frac{\Gamma}{J})}{2\!\:J\!\:\sqrt{1-(\frac{\Gamma}{J})^2}}$ \cite{bender2007faster,zheng2013observation}. The ``non-Hermitian'' means the gate is realized by evolving under a non-Hermitian Hamiltonian.  When $J<\Gamma$, the density matrix will rapidly converge to a steady value \cite{ding2021experimental}, and the error between $\rho_{\textrm{id}}$ and $\rho_m(\tau)$ will be bounded, thus in order to demonstrate the performance of the protection scheme, we mainly focus on $J>\Gamma$ in the following simulation. In the first three subsections, the protection scheme is based on the CPMG-like sequence as introduced in Sec. \ref{cpmglike}. In the last subsection, we demonstrate results obtained from the scheme based on the CPMG sequence (Sec. \ref{cpmg}).
In the following simulations, the DD pulses are assumed to be perfect. $|\psi(t=0)\rangle=|0\rangle$ or $|1\rangle$. The cycle number $m$ is $2$, $4$ or $8$. The noises are generated from Gaussian or uniform distributions. In one round of the simulation, the noise is piecewise constant with period $p$, and we set $p=2 \tau$, although the noise and the sequence have same periods, they are not synchronized with each other, i.e., the noises might undergo a sudden change during the $\tau$ intervals of the sequence.  The simulation is repeated $10000$ times with the noises drawn from the same distribution, and the final density matrix $\rho_m(\tau)$ is obtained by averaging \cite{ding2021experimental}. The fidelities are calculated by \eqref{fid}. We vary $J$, $\tau$, parameters of the noise distributions, and plot the corresponding fidelities. 

\subsection{Constant detuning error}\label{consdet}
\begin{figure}[htb]  
	\makeatletter
	\def\@captype{figure}
	\makeatother
	\includegraphics[scale=1]{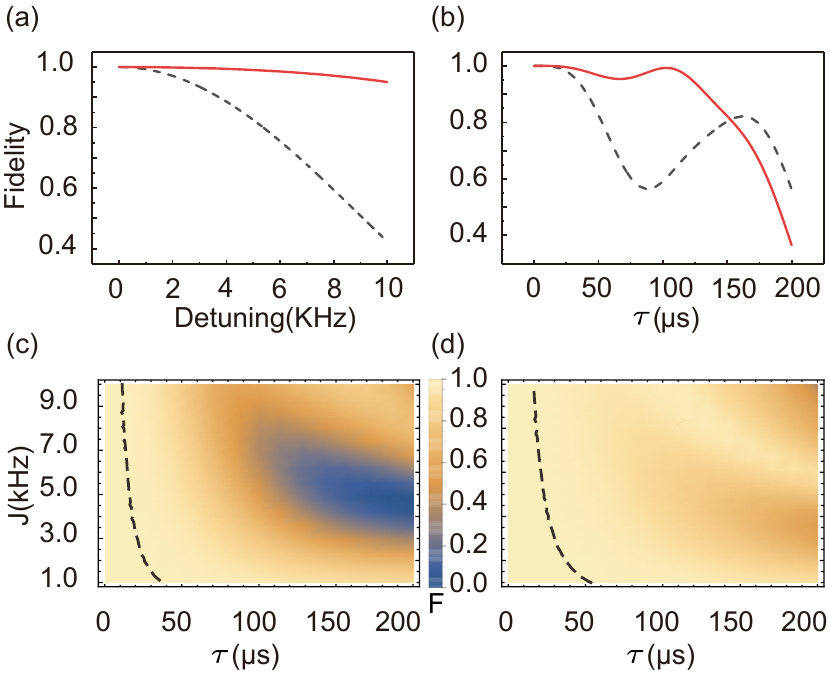}
	\caption{(a) Fidelity of the unprotected non-Hermitian NOT-gate (black, dashed) and the protected one (red, solid) versus detuning error $\beta$. Here $m=2$, $J=10$ kHz,  $\Gamma=1$ kHz, $|\psi(t=0)\rangle=|1\rangle$, $\tau=73.9$ $\mu s$. (b) Fidelity of the unprotected evolution $U(H_{\textrm{PT}},2 \tau)$  (black, dashed) and the protected one (red, solid) versus $\tau$. Here $m=2$, $J=10$ kHz, $\Gamma=1$ kHz, $|\psi(t=0)\rangle=|0\rangle$ and $\beta=2000\pi$ Hz. (c,d) Fidelity of the unprotected evolution $U(H_{\textrm{PT}},2 \tau)$ (c) and the protected one (d) with $m=2$, $J\in [1,10]$ kHz, $\Gamma=1$ kHz, $|\psi(t=0)\rangle=|0\rangle$, $\tau \in [0,200]$ $\mu$s, and $\beta=2000\pi$ Hz. The dashed black line represents the contour with $F=0.999$ (in the region $\tau < 100$    $\mu$s).}
	\label{fig1}
\end{figure}

First we consider constant detuning error, i.e., $\beta$ is a constant. The detuning error could be caused by, e.g., a miscalibration of the quantization magnetic field. $\delta_{\Gamma}=0$ in this subsection. In Fig. \ref{fig1}, (a), the 
performance of $U_{\textrm{NOT}}$ with and without DD protection is illustrated.  
Here 
$m=2$, $J=10$ kHz, $\Gamma=1$ kHz, $|\psi(t=0)\rangle=|1\rangle$, $\tau=T_{\textrm{NOT}}/2=73.9$ $\mu$s.
From the simulation results, we see that the fidelity of the DD protected gate remains high, while the fidelity of the unprotected gate drops quickly as the detuning error is increased. 

To show that our method is also capable of protecting an arbitrary evolution, we choose $|\psi(t=0)\rangle=|0\rangle$ and vary $\tau$ of the sequence, with fixed $\beta=2000 \pi$ Hz, $m=2$, $J=10$ kHz, $\Gamma=1$ kHz, to realize arbitrary evolutions
$U(\tilde{H}_{\textrm{PT}},T=2\tau)$.
The result is shown in Fig. \ref{fig1}, (b). The fidelity of the DD protected evolution remains close to $1$ when $\tau$ is small, in contrast to the unprotected evolution. The fidelity of the DD protected evolution eventually drops as $\tau$ gets larger.
The oscillations of the fidelities are related to the fact that all the elements of $\rho_{\textrm{id}}(T)$ are evolving with a period $T_p=\pi/\sqrt{J^2-\Gamma^2}\approx320  \mu$s, i.e.,  $\rho_{\textrm{id}}(T)=\rho_{\textrm{id}}(T+kT_p )$($k$ is an integer) when $\Gamma<J$.

In Fig. \ref{fig1}, (c), (d), we fix $m=2$, $\Gamma=1$ kHz, $|\psi(t=0)\rangle=|0\rangle$, $\beta=2000 \pi$ Hz, and vary $\tau \in [0,200]$ $\mu$s and $J\in [1,10]$ kHz. The dashed black line represents the contour with $F=0.999$ (in the region $\tau < 100$    $\mu$s). It is clear that the protected evolution has good performance (e.g., $F>0.999$) in a wider range of $\tau$ and $J$ than the unprotected one.

\subsection{Time-varying detuning noise}
\begin{figure}[htb]  
	\makeatletter
	\def\@captype{figure}
	\makeatother
	\includegraphics[scale=1]{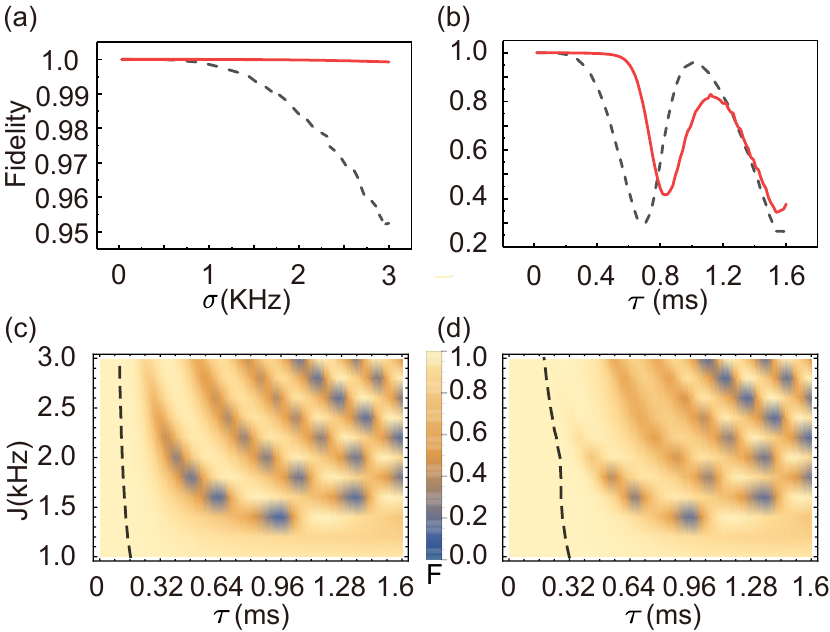}
	\caption{(a) Fidelity of the unprotected non-Hermitian NOT-gate (black, dashed) and the protected one (red, solid) versus standard deviation $\sigma$ of the detuning noise. Here  $m=8$, $J=1$ kHz, $\Gamma=0.5$ kHz, $|\psi(t=0)\rangle=|1\rangle$, $\tau=151.2$ $\mu s$. (b) Fidelity of the unprotected evolution $U(H_{\textrm{PT}},4\tau)$  (black, dashed) and the protected one (red, solid) versus $\tau$. Here  $m=4$, $J=1$ kHz, $\Gamma=0.5$ kHz, $|\psi(t=0)\rangle=|0\rangle$, $\sigma=1.2$ kHz. (c,d) Fidelity of the unprotected evolution $U(H_{\textrm{PT}},4 \tau)$ (c) and the protected one (d) with $m=4$, $J\in [1,3]$ kHz, $\Gamma=1.2$ kHz, $|\psi(t=0)\rangle=|0\rangle$, $\tau \in [0,1.6]$ ms, and $\sigma=1.2$ kHz. The dashed black line represents the contour with $F=0.999$ (in the region $\tau < 0.64$    ms).  }
	\label{fig2}
\end{figure}

Next, we consider time-varying detuning noise. The noise is modeled as a Gaussian distribution with zero mean and standard deviation $\sigma$. It is piecewise constant with period $p=2 \tau$ (it is not synchronized with the DD pulses  as explained above Sec. \ref{consdet}). To calculate the fidelity, the whole process is repeated $10000$ times, and the  final density matrix is obtained by averaging. $\delta_{\Gamma}=0$ in this subsection. In Fig. \ref{fig2}, (a), $m=8$, $J=1$ kHz, $\Gamma=0.5$ kHz, $|\psi(t=0)\rangle=|1\rangle$, $\tau=T_{\textrm{NOT}}/8=151.2$ $\mu s$, and the fidelity of $U_{\textrm{NOT}}$ is plotted as $\sigma$ is varied. It is clear that the performance of the protected gate is better than the unprotected one. 
In Fig. \ref{fig2}, (b), $m=4$, $J=1$ kHz, $\Gamma=0.5$ kHz, $|\psi(t=0)\rangle=|0\rangle$, $\sigma=1.2$ kHz, and the fidelity of  $U(\tilde{H}_{\textrm{PT}},T=4\tau)$ versus $\tau$ is plotted. Again, the performance of the protected evolution is better than the unprotected one at small $\tau$. The fidelity of the protected evolution starts to drop as $\tau$ gets larger. Oscillations of the fidelities similar to Fig. \ref{fig1}, (b) are also observed.

In Fig. \ref{fig2}, (c), (d), we fix $m=4$, $\Gamma=1.2$ kHz, $|\psi(t=0)\rangle=|0\rangle$, $\sigma=1.2$ kHz, and vary $\tau \in [0,1.6]$ ms and $J\in [1,3]$ kHz. The dashed black line represents the contour with $F=0.999$ (in the region $\tau < 0.64$    ms). It is clear that the protected evolution has good performance (e.g., $F>0.999$) in a wider range of $\tau$ and $J$ than the unprotected one. When $J<\Gamma$, the fidelities of the unprotected and protected evolutions are both high, as explained above Sec. \ref{consdet}.

\subsection{Time-varying dissipative-beam noise}
\begin{figure}[hbt]  
	\makeatletter
	\def\@captype{figure}
	\makeatother
	\includegraphics[scale=1]{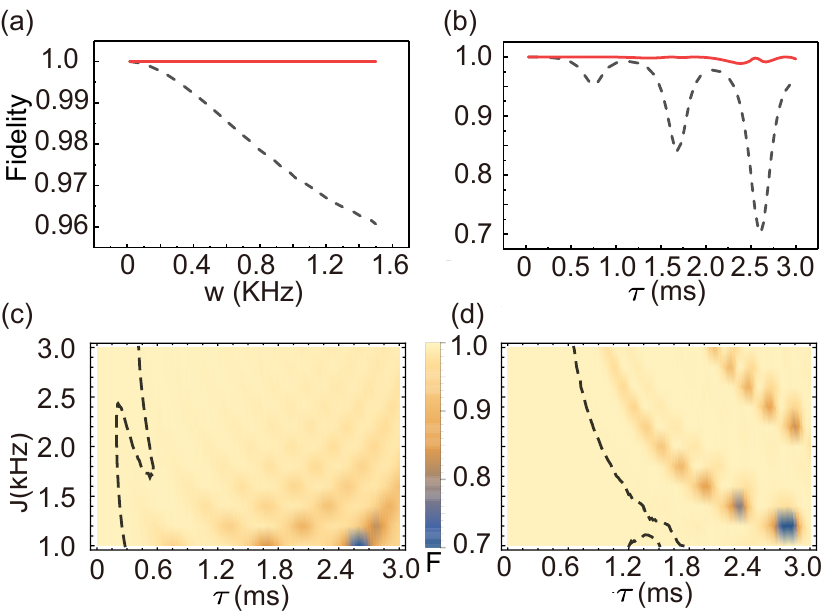}
	\caption{(a) Fidelity of the unprotected non-Hermitian NOT-gate (black, dashed) and the protected one (red, solid) versus the strength $w$ of the dissipative beam noise. Here $m=8$, $J=1$ kHz, $\Gamma=0.5$ kHz, $|\psi(t=0)\rangle=|1\rangle$, $\tau=151.2$ $\mu s$. (b) Fidelity of the unprotected evolution $U(H_{\textrm{PT}},4\tau)$  (black, dashed) and the protected one (red, solid) versus $\tau$. Here $m=4$, $J=1$ kHz, $\Gamma=0.5$ kHz, $|\psi(t=0)\rangle=|0\rangle$ and $\delta _{\varGamma} \in [0,w=100 \, \textrm{Hz}]$. (c,d) Fidelity of the unprotected evolution $U(H_{\textrm{PT}},4 \tau)$ (c) and the protected one (d) with $m=4$, $J\in [1,3]$ kHz, $\Gamma=0.5$ kHz, $|\psi(t=0)\rangle=|0\rangle$, $\tau \in [0,3]$ ms, and $w=100$ kHz. The dashed black line represents the contour with $F=0.999$ (in the region $\tau < 2.4$ ms).}
	\label{fig3}
\end{figure}

Next, we consider the time-varying dissipative-beam noise, which is genuine to evolutions under non-Hermitian Hamiltonians. The noise $\delta _{\Gamma}$ could be caused by e.g., the stray light along the dissipative beam path.  In more complicated experiments where multiple dissipative beams are needed \cite{barreiro2011open}, $\delta _{\Gamma}$ could also be caused by imperfections in other dissipative beams that are not intended to affect the population on $|1\rangle$. $\delta _{\Gamma}$ is modeled to obey random uniform distribution in the range $[0,w]$. It is piecewise constant with period $p=2 \tau$ (it is not synchronized with the DD pulses  as explained above Sec. \ref{consdet}). The whole process is repeated $10000$ times and the density matrix is obtained by averaging. $\beta=0$ in this subsection. In Fig. \ref{fig3}, (a), we choose $m=8$, $J=1$ kHz, $\Gamma=0.5$ kHz, $|\psi(t=0)\rangle=|1\rangle$, $\tau=T_{\textrm{NOT}}/8=151.2$ $\mu s$. The fidelity of the non-Hermitian NOT gate is plotted against the strength of the dissipative-beam noise $w$, and the fidelity of the DD protected gate is higher than the unprotected one. In Fig. \ref{fig3}, (b), $m=4$, $J=1$ kHz, $\Gamma=0.5$ kHz, $|\psi(t=0)\rangle=|0\rangle$, $w=100$ Hz, and the fidelity of $U(H_{\textrm{PT}},4\tau)$ is plotted with increasing $\tau$. The performance of the protected evolution is better than the unprotected one. Oscillations of the fidelities similar to Fig. \ref{fig1}, (b) are also observed.

In Fig. \ref{fig3}, (c), (d), we fix $m=4$, $\Gamma=0.5$ kHz, $|\psi(t=0)\rangle=|0\rangle$, $w=100$ Hz, and vary $\tau \in [0,3]$ ms and $J\in [1,3]$ kHz. The dashed black line represents the contour with $F=0.999$ (in the region $\tau < 2.4$ ms). The protected evolution has good performance (e.g., $F>0.999$) in a wider range of $\tau$ and $J$ than the unprotected one.

\subsection{Protection based on the CPMG sequence }
\begin{figure}[hbt]  
	\makeatletter
	\def\@captype{figure}
	\makeatother
	\includegraphics[scale=1]{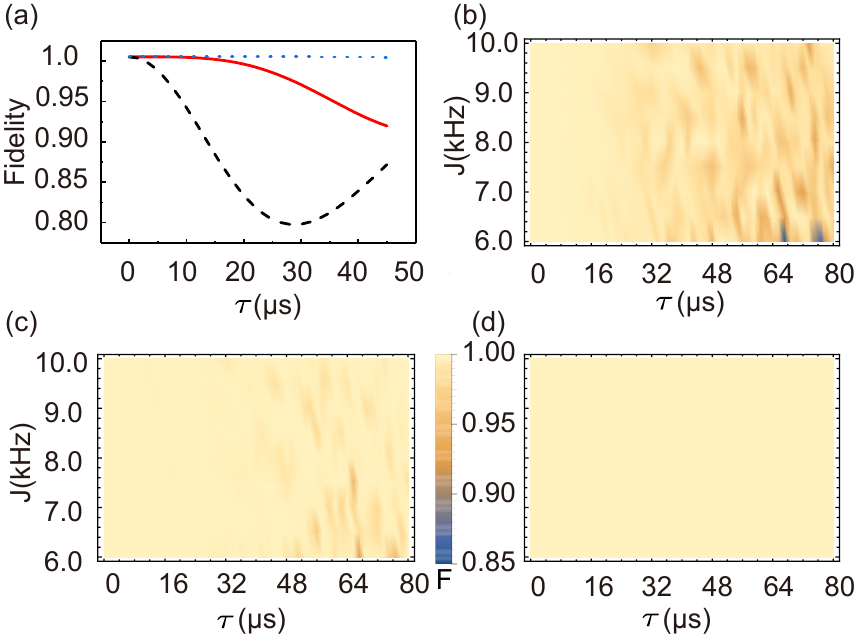}
	\caption{(a) Fidelity of the unprotected evolution (black, dashed), the protected one ($s_1$) based on the CPMG-like sequence (red, solid), and the protected one ($s_2$) based on the CPMG sequence (blue, dotted) versus $\tau$. Here $m=4$, $J=10$ kHz, $\Gamma=1$ kHz, $|\psi(t=0)\rangle=1/\sqrt{2}(|0\rangle+|1\rangle)$, $\beta=2000 \pi$ Hz and $\delta_{\Gamma}=2000$ Hz. (b,c,d) Fidelity of the unprotected evolution $U(H_{\textrm{PT}},4\tau)$ (b), the protected one ($s_1$) based on the CPMG-like sequence (c), and the protected one ($s_2$) based on the CPMG sequence (d). Here $m=4$, $J\in [6,10]$ kHz, $\Gamma=1$ kHz, $|\psi(t=0)\rangle=1/\sqrt{2}(|0\rangle+|1\rangle)$, $\tau \in [0,80]$ $\mu$s, $\sigma=1200$ Hz, and $w=100$ kHz. }
	\label{fig4}
\end{figure}

Finally, we further present the performance of protection based on the CPMG sequence ($s_2$) discussed in Sec. \ref{cpmg}, which achieves second-order protection, and compare it with the unprotected evolution and the CPMG-like sequence ($s_1$). In Fig.\ref{fig4}, (a), $m=4$, $J=10$ kHz, $\Gamma=1$ kHz. To show the generality of our scheme, we consider simultaneous existence of constant errors $\beta=2000 \pi$ Hz and $\delta_{\Gamma}=2000$ Hz, and set $|\psi(t=0)\rangle=1/\sqrt{2}(|0\rangle+|1\rangle)$.  The fidelity of the evolution under $s_2$ is above $s_1$ and the unprotected evolution. 

In Fig.\ref{fig4}, (b,c,d), $m=4$, $J \in [6,10]$ kHz, $\Gamma=1$ kHz, $|\psi(t=0)\rangle=1/\sqrt{2}(|0\rangle+|1\rangle)$, and $\tau \in [0,80]$ $\mu$s. To further demonstrate the generality of our scheme, we still consider simultaneous existence of constant errors $\beta$ and $\delta_{\Gamma}$, while this time we pick $\beta$ from a Gaussian distribution with standard deviation $\sigma=1200$ Hz, and $\delta_{\Gamma}$ from a uniform distribution in the range $[0,w=100]$ Hz in each repetition of the simulation. The simulation is repeated $10000$ times (at each $(\tau,J)$), and the density matrix is obtained by averaging. 
The fidelities of $s_2$ and $s_1$ are higher than the unprotected evolution in a wide range of $\tau$ and $J$ ($F>0.93$ for $s_1$ and $F>0.999$ for $s_2$ in the chosen range). Thanks to the ability to reach second-order protection, the fidelity of $s_2$ is also higher than $s_1$ in a wide range of $\tau$ and $J$.

 From these simulation results, it is clear that the DD- protected non-Hermitian evolutions have better performance than unprotected ones under noises acting along the quantization axis, indicating the effectiveness of our method.

\section{Conclusion}\label{Conclusion}
In summary, we achieve the protection of non-Hermitian $\mathcal{PT}$-symmetric Hamiltonians against
noise acting along the qubit's quantization axis by combining quantum evolutions with dynamical decoupling sequences.
We demonstrate the performance of our method by numerical simulations. We choose realistic noise sources and parameters including: constant detuning error, time-varying detuning noise and dissipative beam noise. The fidelities of the protected evolutions are well above the unprotected ones under all the above situations, demonstrating the effectiveness of our method.
We thus conclude that we have theoretically demonstrated the protection of $\mathcal{PT}$-symmetric Hamiltonians against noise and errors. Experiments using this method in a trapped ion quantum processor is being implemented in our lab. Our work paves the way for further studies and applications of non-Hermitian $\mathcal{PT}$-symmetric physics in noisy environment.

\section*{Acknowlegments}
We thank the reviewer for helpful comments. Support come from the Key-Area Research and Development Program of Guangdong Province under Grant
No. 2019B030330001, the National Natural Science
Foundation of China under Grant No. 11774436, No.
11974434 and No. 12074439, the fundamental research
funds for the Central Universities (Sun Yat-sen University, 2021qntd28), the Central-leading-local Scientific and Technological Development Foundation under Grant No. 2021Szvup172. Le Luo receives support from Guangdong Province Youth Talent Program under Grant No.
 2017GC010656. Yang Liu receives support from
Natural Science Foundation of Guangdong Province under Grant No. 2020A1515011159, Science and Technology program of Guangzhou under Grant No. 202102080380. Ji Bian receives support
from China Postdoctoral Science Foundation under Grant No. 
2021M703768.
\bibliographystyle{unsrt}
\bibliography{dynamical_decoupling}

\end{document}